# "I got in trouble": A case study of faculty "doing school" during professional development

Alice Olmstead[1] and Chandra Turpen[2]
[1]University of Maryland, Department of Astronomy,
1113 Physical Sciences Complex, College Park, MD 20742
[2]University of Maryland, Department of Physics,
Toll Physics Building, College Park, MD 20742

**Abstract.** Professional development workshops are commonly used to promote the adoption of research-based instructional strategies among physics and astronomy faculty. After learning about such strategies, faculty are often motivated to modify and adapt them within their own classrooms, but prior research shows they may be underprepared to do so in ways likely to maintain the positive student outcomes the designers were able to foster. In this paper, we analyze the experiences of a focal group of faculty during one session of the Physics and Astronomy New Faculty Workshop, where they are asked to engage in a task as mock physics students. We compare their experiences to student behaviors documented in others' research, and find that their group coordination and sense-making poorly represent the kinds of interactions our community would encourage them to foster in their own students. We briefly discuss the implications of these preliminary findings for professional development and our plans for future research.

## I. INTRODUCTION

Our community has made significant progress in understanding how undergraduate students learn physics and astronomy, and has developed a range of research-based and research-validated instructional strategies (RBIS) that can help faculty to facilitate meaningful student participation in their courses [1]. Many faculty become motivated to try and do try RBIS, but the results have not been as transformative for their instruction nor as sustained as one might hope [2]. The capacity of physics and astronomy education research to improve student outcomes seems to break down at the stage where our community expects faculty to take up RBIS exactly as presented to them—an expectation that goes against what many faculty want or need [3]. Instead, faculty often implement RBIS differently than the designers, sometimes in ways significantly less aligned with education research principles than the original implementations [3]. Because of this, they likely need more support and guidance in learning why a particular RBIS implementation might be successful and deciding when adaptations or modifications seem appropriate given their local contexts. These challenges motivate critical examination into how faculty learn about teaching innovations when they participate in professional development (PD) activities.

## II. THE NEW FACULTY WORKSHOP

Our research focuses on the Physics and Astronomy New Faculty Workshop (NFW): a national workshop that has played a central role in increasing faculty's awareness of and experimentation with RBIS over the past 19 years [2]. About 70 faculty attend each workshop, which accounts for about half of new tenure-track physics hires at 4-year institutions each year. The 4-day workshop is split into 45 to 60-minute sessions led by experienced physics and astronomy education researchers and instructors, and a majority of the sessions highlight specific RBIS.

The NFW offers a rare opportunity for faculty to experience RBIS implementation, which could serve as a concrete anchor for reimagining future instruction. We believe that the more faculty are encouraged to reason about potential affordances and drawbacks of a variety of instructional choices within scaffolded activities, the more likely they will be to move towards highly desirable teaching practices. While post-workshop surveys indicate that faculty find many aspects of the NFW to be highly useful, it is difficult to distinguish increased awareness and motivation from an increased ability to implement and assess teaching decisions without evidence of faculty's thinking and behaviors in situ. Therefore, we set out to address the following research question by observing the workshop itself: *How could faculty's experiences at the NFW improve their ability to enact, evaluate, and/or adapt RBIS?*

## III. METHODOLOGY

We videotaped three iterations of the NFW, and the first author coded about half of the teaching-focused sessions from one iteration (partly during the workshop, partly from video) using a workshop observation tool that we are currently developing. Two sets of codes comprise the tool: one describes the form of faculty's engagement and is modeled after existing classroom observation tools [4,5]; the other describes the focus of faculty's engagement and draws from successful PD practices reported in the K-12 literature [6]. Ultimately, data from our tool will both allow PD leaders to reflect on the alignment between workshop design, workshop goals, and effective learning activities for teachers, and allow researchers to select and analyze excerpts that contain significant faculty discourse and actions.

Here, we use our tool for the second purpose, and identify a 9-minute period of small group discussion in which faculty are positioned as physics students while a workshop leader (WL) simulates the instructor's role during the implementation of a RBIS [6]. We selected this segment because it was the longest period of small group discussion within the NFW sessions we had coded; a majority of sessions, including this one, are primarily lecture-based. Although the contents of the workshop leaders' lectures may be valuable, we do not expect to find concrete evidence of faculty's thinking during lecture, nor would we expect faculty to substantially improve their ability to reason about teaching without some active engagement. We think that collaborative, student-like interactions could provide a valuable learning opportunity for faculty, and act as a mechanism by which they improve their understanding of how to facilitate group work. Faculty's facilitation skills will strongly influence their ability to implement RBIS: collaboration is central to most RBIS [3], and not all collaboration is equally beneficial to students [7,8]. We unpack how faculty behave as pseudo-students in order to understand how they might improve from reflecting on this experience.

A single focal group of faculty was recorded during each NFW session. Here, we focus on three key episodes featuring the four focal group members in this session, given the pseudonyms Ted, Maggie, Rachel, and Brad. We use Barron (2000)'s markers of coordination in group work [9], as well as studies of how students feel they need to act to be successful in school [10-12], to compare their experiences to those of students. We find that although these faculty appear genuinely immersed in enacting student roles, their behaviors do not exemplify cooperative, equitable, or intrinsically motivated student behaviors that we would want them to bring out in their students. Specifically, as we describe below, many of their interactions are consistent with markers of low coordination with their peers, and they focus more on "doing school" than "doing science".

## IV. FACULTY ACT AS STUDENTS

Before the first episode begins, all four faculty are seated and read instructions projected at the front of the room that will lead them through an activity about conceptualizing plane waves. The first step directs them to draw a grid on a large whiteboard lying on the table between them, with at least 7 x 7 points spaced approximately 2 inches apart. Maggie reads aloud softly. After a few seconds, before others appear to be finished reading, Ted stands up, takes the cap off a marker and leans forward as if to draw on the whiteboard. He glances back at the instructions as the WL begins to speak.

*Episode 1: The whiteboard as contested territory*

WL: Alright. \\You have one minute to get those dots up there. Make them as square as you can in one minute.][1]
Maggie: \\Before we draw, why don't we actually measure it?]
Ted: But I mean approximately two inches (motions as if to start drawing)
Rachel: Those are gonna be a centimeter right? (pushes a piece of paper onto the whiteboard, blocking Ted)
Ted: I think approx-, I mean
Maggie: Listen we wanna be accurate,
Ted: \\Okay.]
Rachel: \\No no but at least you have a straight line] (pushes the paper towards Ted)
Maggie: \\2.54cm.]
Rachel: You have a straight line.
Ted: You do it. (shrugs and pushes the paper back towards Rachel)

Maggie, Rachel, and Ted take up competing aspects of the WL's instructions: Maggie and Rachel attempt to be highly accurate, which aligns with an interpretation that the squareness of the grid is important, while Ted makes several bids to draw the grid "approximately" and repeatedly motions as if to start drawing, which aligns with the directive to draw the grid quickly, in "one minute." They do not offer any justification for their arguments, which may suggest that correctly interpreting the WL's rules takes priority over deciding what level of accuracy is appropriate for the task, consistent with students "doing school" [10]. The way that faculty physically and verbally negotiate who will draw on the whiteboard and how this drawing will be done puts the whiteboard in the center of their conflict. Treating a group artifact as contested "territory" is a marker of low coordination in group work, as are "conflicts of insistence" (their conflict does not build meaning), and violation of turn-taking norms (faculty repeatedly interrupt each other and talk simultaneously) [9].

Immediately after Episode 1, the WL comes over to their group, pushes the paper off the whiteboard, and starts drawing a grid on it. She does not question faculty about what they were doing previously and her actions functionally discard Rachel's approach. Ted vies for the WL's approval of his idea, claiming "That's what I was gonna do until… *(pointing towards his group)*". When she walks away, Ted asserts to his peers "I was about to do that very same thing until I got in trouble." Consistent with our initial claim, this statement and Ted's interaction with the WL also imply a "doing school" mentality: Ted is trying to ap-pease an authority figure and to establish himself as a "good student" set apart from his peers [11].

---

[1] "// ]" notation indicates simultaneous speech.

As Ted complains about getting in trouble, he, Brad, and Maggie begin drawing points on the whiteboard. Ted is the only group member who is standing and draws points across the whole whiteboard without pause; Maggie and Brad only draw points near the corners and edges of the grid and do so intermittently. Twice, Brad pulls away when he and Ted try to draw a point at the same location on the grid. The second time, Ted laughs and remarks, "How many physicists does it take to screw in a light bulb?", while Rachel, who is watching, jokes that the grid is made up of "drunken points." As Maggie finishes drawing and pulls away, Ted reaches across the table and adds two more points directly in front of her. Ted stands up and re-caps his marker, and Episode 2 begins.

*Episode 2: Maggie uses gender for role negotiation*

Maggie: This is why men don't get to draw things. No that's 8. Oh that's 8. You should let the women draw it.
Ted: (Laughs) \\Well it's a good thing I work on non-Euclidean geometry.]
Rachel: \\It's hung like men hang wallpaper]
(Maggie says something inaudible. She erases and re-draws several points.)
Rachel: Although to be fair we shouldn't say things \\like this],
Brad: \\I tried.] (Smiles and shakes his head.)
Rachel: because if \\they said this is drawing like women drive then we'll get in trouble.]
Ted: \\Yeah. Oh my god.]

Episode 2 reinforces our earlier claim that the group members compete for use of the whiteboard, treating it as territory. Maggie tries to re-negotiate her role both within the task and relative to the whiteboard, using gender to position herself and Rachel as more competent at drawing the grid than Ted and Brad. From an equity standpoint, it seems consequential that Maggie promotes her own participation by assigning herself a drawing task, when secretarial roles are more often implicitly or explicitly assigned to female students than male students and can limit female students' access to learning during group work [8]. Her comments also suggest that gender plays a significant role in how she perceives their unequal participation in the task, and may be indicative of a larger underlying tension throughout these episodes. Still, this discursive move opens up a more active role for her within the task up to this point, without explicitly pointing out that Ted has been allocating most of that responsibility to himself. Ted, Rachel, and Brad all smile or laugh, reacting as if it was a joke, and Maggie successfully takes on the role that she made accessible to herself, thus temporarily gaining control of the whiteboard [13]. Rachel sustains an earlier aspect of "doing school" by revoicing Ted's phrase about getting in trouble, now providing it as a potential risk of making comments about gender stereotypes. She seems to perceive that an aversion to breaking the rules of the classroom will be a valid motivator within her group, thus assuming they have a shared, school-like desire to win the favor of the WL [11].

After completing the construction of the grid, the group responds to the prompt: "For every point on your grid…connect the points with equal values of $\boldsymbol{k} \cdot \boldsymbol{r}$", where the WL has introduced a different vector $\boldsymbol{k}$ to each group and $\boldsymbol{r}$ is the position vector from the origin. For this group, $\boldsymbol{k}$ is the vector with components 1 and 2, which makes the solution lines for which $x + 2y = c$, where $c$ is a constant. Maggie states, "I don't understand what she's asking" and re-reads part of the instructions aloud. Ted begins to articulate some ideas that will go into the solution, such as "$k$ dot $r$ is just an equation so it's just lines", but leaves many of his sentences unfinished. Just before Episode 3 starts, the WL pauses all the participants and gives them additional guidance about what to do next.

*Episode 3: Low coordination of mathematical reasoning*

Ted: So this, so $k$ is, $k$ is (1, 2). Right? That's a vector. (writing across the whiteboard facing himself)
Maggie: Wait why are you? Why //don't you just draw it $\hat{x} + 2\hat{y}$?] (also writing on the whiteboard, on the nearest corner to her)
Ted: //Just, we're doing an $x$ component. So yeah, so $k$ dot $x$, $k$ dot $r$] is gonna be equal to $x + 2y$. Right? (Maggie nods.) You with me?
Brad: Yeah.

Lending weight to our claims from the previous episodes, even as faculty progress to a more challenging part of the task, low coordination persists. The whiteboard still seems to be perceived as territory, once again primarily controlled by Ted: he writes equations near the center of the board, upside down to everyone but himself. Maggie asks "why" and her tone indicates that she is frustrated. Violation of turn-taking norms continue and now are more consequential towards developing a shared understanding of the solution. Although Ted is likely aware that he and Maggie speak simultaneously, he does not acknowledge or attempt to repair this social misstep even though she suggests a viable alternative representation. Ted looks at Maggie frequently and seeks signs of confirmation that she is listening to him, but neither responds to her proposal nor articulates all parts of his thinking, discarding some of his own ideas without pause or explanation [9]. In these ways, Ted launches into constructing the solution independently as the session continues.

## V. DISCUSSION

It is a complex undertaking for an instructor to create learning environments in which their students regularly elicit, listen to, and build on each other's ideas while focusing on the pursuit of scientific meaning. Although asking faculty to engage in collaborative learning tasks at workshops could provide authentic versions of the kinds of experiences we want faculty to create for their students, our data reveal that faculty's actual experiences may look quite different from this ideal. While other groups may have fared better, it is clear that some faculty will struggle to collaborate in workshop settings, and that a workshop leader's facilitation moves can reinforce faculty's sense that they are "doing school" as opposed to doing physics.

While we do not know the goals of the workshop leader surrounding this particular task, we argue that these faculty's behaviors and actions underscore the importance of teaching faculty to notice and address problematic student interactions such as unequal participation, a lack of attentiveness to others' ideas, and a focus on a standard of achievement that carries little weight outside of school contexts. Though faculty will rarely find themselves taking on student roles in school-like environments, they might frequently find themselves teaching students who routinely act in these problematic ways in the classroom. If a workshop leader chose to centrally pursue this goal and was able to allocate time to unpacking the task, experiences like the ones described here could be generative. For example, a workshop leader could guide faculty to reflect on what facilitation moves supported or inhibited collaboration in their group, how they might want their students' experiences to differ from their own, and what a facilitator could have done differently to promote these shifts in student engagement. Similarly, video or case study examples could be used to promote discussion and provide alternatives scenarios for faculty to compare to what they experienced [5,14]. These examples could narrow in on specific aspects of group interaction like gender dynamics, which might be difficult to discuss otherwise because of existing social tensions.

We can also envision other workshop session goals that could be better met through modifications to the implementation of this task. If the workshop leader had been able to spend more time listening to faculty's ideas and intervening in ways that promote collaboration and sense-making [15], faculty might be more motivated to use group work in their own classrooms, while highly attentive faculty might notice and adopt these facilitation moves. Alternatively, a workshop leader might choose to target faculty's pedagogical content knowledge by framing, facilitating, and/or debriefing the task in ways that encourage faculty to consider how it would play out given students' evolving disciplinary knowledge.

Finally, we note that these results are preliminary and do not represent a complete picture of faculty's engagement within the NFW. More work is needed to understand how other faculty engage in similar activities, and what other PD practices might contribute positively to faculty's learning about teaching.

## ACKNOWLEDGEMENTS

This work is supported by funding from NSF DUE-1431681. The authors also thank Andy Elby, Vijay Kaul, Erin Ronayne Sohr, Deborah Hemingway, and Kim Moore for feedback on this analysis; Natasha Holmes, Angie Little, Ida Rodriguez, and Paula Heron for their reviews at the 2015 FFPER graduate student/postdoc symposium; Ed Prather and Derek Richardson for their support and guidance; and Bob Hilborn and the workshop leader in this study for allowing and encouraging us to do this research.